\documentstyle[aps,twocolumn,prb,psfig]{revtex}

\def\pmb#1{\setbox0=\hbox{$#1$}
  \kern-.25em\copy0\kern-\wd0
  \kern.05em\copy0\kern-\wd0
  \kern-.025em\raise.0433em\box0 }

\begin{document}
\draft
\widetext
\twocolumn[\hsize\textwidth\columnwidth\hsize\csname @twocolumnfalse\endcsname

\title { Immunization and Aging: a Learning Process in the Immune Network}
\author{Rita Maria Zorzenon dos Santos}
\address{Instituto de F\'\i sica - Universidade Federal Fluminense, Av.
Litor\^anea, s/n - Boa Viagem, 24210-340,
Niter\'oi-RJ, Brazil}

\author{Am\'erico T. Bernardes}
\address{ Departamento de F\'{\i}sica, Universidade
Federal de Ouro Preto, Campus do Morro do Cruzeiro,
35400-000, Ouro Preto-MG, Brazil}
\date{\today}
\maketitle

\begin{abstract}
The immune system can be thought as a complex network of
different interacting elements.  A cellular automaton, defined in
shape-space, was recently shown to exhibit self-regulation and
complex behavior and is, therefore, a good candidate to model
the  immune system. Using this model to simulate a real immune system
 we find good agreement with recent experiments on mice.
The model exhibits the experimentally observed refractory behavior of the
immune system under multiple antigen presentations as well as
loss of its plasticity caused by  aging.

\end{abstract}
\pacs{PACS: 02.70Lq,05.45+b,87.10+e,87.22}
\vskip2pc]

The special capability of the immune system is pattern
recognition and its assignment is to patrol the body and guard its identity.
The main cells responsible for the immune response against  any
attack by antigens (virus, poison, etc.) are the white blood
cells called lymphocytes. Each lymphocyte carries on its surface the
order of $10^5$ molecular receptors and the immune system is able to produce
the order of $10^{11}$ different types of molecular receptors.
One of the main classes of lymphocytes are the B cells which produce antibodies
with the same molecular shape as those of its molecular receptors.
Each molecular receptor (or antibody) is able to recognize  and
to be recognized. This immune recognition  can lead either
to a positive response, which generates cell proliferation,
cell activation  and antibody secretion, or to a negative one that results in
tolerance or suppression.

In 1974  Niels Kay Jerne \cite{Jerne} proposed a theory which compares the
 immune system to a self-regulated multiconnected
functional network. This functionally
connected network, generated by the (lock-key) interactions  among the
elements of the immune system, is self-regulated by  the dynamics of activation
and suppression; some populations decay and new ones grow,
but there is no percolation of the information through the
entire immune system. The immune memory may be a 
consequence of network interactions. Some experiments \cite{coutinho,holmberg}
suggest that only $20\%$ of the lymphocytes are connected,
whereas their majority are small resting cells, able to
respond to any antigen.

We have simulated recently \cite{bz97} a modified version \cite{zb95} of a
cellular automaton model proposed by Stauffer and Weisbuch \cite{sw92}, which,
in turn, was based on one introduced by De Boer, Segel and Perelson
 \cite{bsp92}. Although the previous studies have shown that this model can
exhibit stable and chaotic behaviors, none of them are interesting from the
immunological point of view. Our simulations
 have demonstrated that in the transition region, between those behaviors, we
obtain  a multiconnected functional network whose emergent properties reproduces
many of the properties of the temporal evolution of the immune repertoire
\cite{bz97}.

The elements of the automaton are associated with points in shape-space
\cite{ps79}. Each point on a  $d$-dimensional space corresponds
to the shape of a molecule (associated with a B cell receptor or antibody).
Every one of the $d$ coordinates describes one of many factors
involved in the immune pattern recognition, such as geometrical shape, 
electric charge, hydrophobicity etc. To each site $i$ we associate an
automaton of three states, representing the population
(concentration) of the corresponding receptor: low - or non-activated
($B_i=0$), intermediate ($B_i=1$) and high - or excited ($B_i=2$). Site $i$
interacts with $2d+1$ sites: its mirror image $-i$ and the nearest
neighbors of the
mirror image (representing imperfect lock-key interactions). For each site
we define a field $h_i$ that represents the concentrations of
complementary receptors,
\begin{equation}
h_i=\sum_{j(i)} B_{j(i)}
\label{eq:field}
\end{equation}
where the sum runs over
the $2d+1$ sites $j(i)$ that influence the
site $i$. The update rule of the automaton is based on a
dose response function which describes the receptor crosslinking involved
in B cell activation \cite{zb95,sw92,bsp92}. There is a minimal dose
($\theta_1$) of antigen/or antibody excitation
that elicits the specific response, but for a high dose of excitation
(greater than $\theta_2$) the response decreases (suppression). The rule is
\begin{equation}
B_i(t+1)= \left\{
\begin{array}{ll}
B_i(t) +1   \qquad  & {\rm if} \qquad \theta_1 \leq h_i(t) \leq \theta_2 \\
B_i(t) - 1  &{\rm otherwise}
\end{array}
\right.
\label{eq:rule}
\end{equation}
but no change is made if it would lead to $B_i=-1$ or $B_i=3$.

This model exhibits a transition from stable to chaotic regimes for
$d\geq 2$ \cite{zb95}, depending on the activation threshold
($\theta_1$) and the width of the activation interval of the window
($\theta_2 - \theta_1$). In the stable region the system always evolves to
fixed points or short limit cycles, with very low percentage of activated sites,
 whereas in the chaotic region most sites are excited, corresponding to
a non-healthy state. None of these behaviors corresponds to the dynamical
 picture of the functional connected network
suggested by Jerne. In the transition region 
(whose location depends on $d$ \cite{zb95}) we did observe, however,
emergent complex behavior which is appropriate  to describe a self-regulated
multiconnected network \cite{bz97}. We have also shown that  memory comes
 from the dynamics of the system and can be described as the ability of the
system to adapt to changes provoked by excitations (antigen presentations).

We use this model to simulate the main features
observed in some recent experiments
 concerning immunization and the effects of aging on
the immune response of mice \cite{bruno97,lahmann,ana97}. Since until now
there is only limited evidence supporting the existence of the immune
network \cite{coutinho,holmberg} but no conclusive proof, our simulations
may
 provide the much needed connections between the available
experimental results and the Jerne theory.

Verdolin \cite{bruno97} subjected
young (8 weeks old) mice to a  first
immunization, and 14 days afterwards to 9
consecutive immunizations  with intervals of one week
between them. The specific antibodies are measured 7 days after each
antigen presentation. The protocol used for immunization consist of
 intraperitoneal
injection (boosters) of a specific antigen (OVA). 
Verdolin's results (for a group of
6 mice)  show (see Fig 1) the
specific antibody counts as a function of the presentation with all doses
counted relative to the first immunization response. The results show a fast
increase from the first to the  fourth presentation with a consecutive
saturation of the response. This is interpreted from the
immunological point of view as a refractory behavior  of the immune system,
meaning that after few presentations there is a saturation of the level of
antibody counts, a fact that can not be explained only by clonal
selection theory for the immune system.

The second experiment we discuss in this paper concerns the
effects of aging on the immune responsiveness.
According to the results obtained by Lahmann {\it et al.} \cite{lahmann}
and, later, by Faria {\it et al.} \cite{ana97} for young (8 weeks)
and
old (25
 weeks) mice, after the first immunization the response of the young mice
achieves much higher levels (in the antibody counts) than the response
generated by the old ones, indicating some rigidity of the system, acquired
during the aging process. In other words, there is a loss of responsiveness
(see the insert of Fig 1).

We set out to test whether the model (1-2) is able to reproduce the
experimental findings, of saturation and aging
described above. We used the following initial distribution for the $B$
 variables: $(1-x)$ of $B=0$, and ${{1} \over {2}} x $ of $B=1$ and $2$. 
All our simulations were performed at the transition region for a 
three-dimensional lattice with
$L=50$;  the results are qualitatively the same as those obtained with $L=100$
and for higher dimensions \cite{bz97}.
 We chose $x=0.26$ and used the   activation interval according to our
previous studies. In order to study the effects of a given antigen
 presentation we proceed as follows: a system, prepared as described above,
was allowed to evolve according to the rules (1-2); this was used as
 our ''control'' group.
In a replica of the same initial system the antigen presentation is
simulated by introducing ''damage''\cite{jancar94}: randomly chosen
regions of receptor populations in the $B=0$ state were
flipped to the activated state($B=2$) \cite{bz97}. This flip
simulates the exposure to antigen indirectly, by the activation or the
increase of the concentration of specific clones (receptors) that are
able to recognize this antigen. The evolution of the control system
and  the damaged copy
was then compared by measuring the Hamming distance between them,  to study the
effects of the antigen presentation. In the case of one or two antigen
presentations \cite{bz97} the results indicate that under perturbation
 the multiconnected network expands and relaxes in few
time steps, following an aggregation/disaggregation dynamics which adapts the
system to the new conditions, including some sites of the damaged regions in
the next configuration of the functional network.

In this work we discarded in all simulations the initial $1000$
Monte Carlo Steps (MCS)- a step consists of one update of all sites. 
At this point( considered to be equivalent to the ''birth''of the 
immune system) we start to produce at random antigen presentations
of different types and sizes, with different, randomly chosen time
intervals between them. This introduced noise simulates a real immune 
system exposed to many antigens, present in the environment in which the 
system lives and is ingested  in food, etc. We adopted the arbitrary time 
scale of $5$ MCS corresponding to one day.  The noise discussed above was
generated by small damages produced at random: the time interval between two
consecutive antigen presentations can vary from $1$ to $100$ MCS
($1-20$ days); each antigen presentation can correspond to $1$, $2$ or $3$
damages introduced at different regions of non-activated  populations.
These regions  may have different sizes, varying from $1$ to $3$ 
(onion-like) concentric layers 
(containing  $7$, $25$ and $63$
sites respectively) around a central site. The control
system and the  replica that  simulates the immunization process
(multiple antigen presentations) were subjected
to the same noise, during  the entire simulation. The immunization 
protocol was simulated by damaging $6$ layers around a specific site
at  fixed time intervals of one week ($35$ MCS), corresponding to the
peritoneal injection. 
This high-dose antigen presentation clearly differentiates  the specific
antigen presentation from the noise. We
performed these sequences of high dose antigen presentations starting at two
different times, corresponding to two different ages,
$8$ weeks (280 MCS) and $25$ weeks (875 MCS),
in order to study the effects of aging
on the immune responsiveness.

In figure 2 we show the results obtained for a sequence of antigen
 presentations. For each case (different age) our results are averaged over
$10$  statistically independent samples corresponding to $10$ randomly
uncorrelated initial configurations (and different noise).

The specific antibodies counts in the experiments are compared in our
 simulations to the number of different excited populations involved
in the response, which is measured by the time evolution of the total
Hamming distance  between the ``control'' and the excited replica.
We obtained the  same qualitative
behavior as the real immune systems shown in figure 1. After few
presentations the  responses saturate, showing the same refractory behavior as
seen in the experiments. Since in the experiments the specific 
response was measured (corresponding to the antigen that was used), we
have also considered, in our simulations, 
the specific response.  To this end we measured the changes on
the clones belonging to  the damaged region (DR) and its mirror image
(M). Saturation is seen also in 
these regions, as shown in Fig 3, which presents  the
specific changes in DR, M and the total HD for a single sample 
(the same behavior was observed in all samples).

The learning process that may occur in the immune network, due to the 
antigen introduction, results in local changes of the activated clones 
configurations, and memory is due to
the attainment of a new steady state after the antigen challenge
disappears. The learning process that may occur due to multiple antigen 
presentations, leads to the saturation behavior:
 after a few antigen presentations the system learns how
to control the excitation provoked by this antigen and does not need to
improve the response anymore.
This interpretation is supported by  other studies performed for high-dose
antigen presentations without noise, that will be published elsewhere.
By observing the evolution of the excited sites in the DR and M regions,
we saw that after each presentation more sites are
included in the functional network. After that, the response will 
stay the same. The multiple antigen presentation, 
lead to the saturation of the number of new sites incorporated in 
those regions and after that the system will always give the same 
response to this kind of antigen. The aggregation/disaggregation
processes guarantee that the subsequent configurations of the network,
arrived at after each antigen presentation, include the necessary information
about the damage introduced, but still keep only $10-20\%$ of the sites
excited, in agreement with the experimental results.

This conclusion is also reinforced by the observation that,
 the level of saturation and also the
time necessary for the system achieve the saturated state will depend on
the size and number of damages which simulate the antigen presentation.
We have performed the same kind of simulation (immunization procedures) 
for different sizes of the damages; for small damages ($4$ layers) the 
system saturates faster than for big ones ($6$ layers), since it learns 
faster about the small damaged regions.

The data shown in the insert of figure 2
corresponds to the saturation values of the Hamming distance difference between
``control'' and replica as measured
for the two different ages ($8$ and $25$ weeks). From those results we
 conclude that the ``older'' the system is, the more rapidly it saturates
 and the less intense is its response. The older system has been exposed
to a larger number of different antigens, thus more regions have already been
excited, more information has been added to the functional network and  the
system is more rigid to accept new local changes. Old systems saturate
faster than young ones, since less changes are accepted in their dynamics. 
The rigidity of the system, which increases with age,
can be interpreted from the network point of view as a loss of its plasticity.
The younger the system the greater its plasticity.

In this work we have shown that the multiconnected functional network obtained
in the critical region of a simple
cellular automaton model can describe the behavior of a real immune system.
The emergent properties of
the cooperative behavior found in our simulations are in good agreement with the
experimental results exhibited by mice in experiments of immunization and aging.
Our results  indicate that the Jerne theory as implemented in this
model  could explain these  experimental observations and possibly provide 
the much needed connection between this theory and the available
experiments.  The immunization process and
the saturation (refractory behavior) observed in the antibody counts
 are interpreted within
the network point of view as the learning process of the immune system
and the aging effects are interpreted as the loss of ability to produce
new local changes or the loss of plasticity. This work also give support
to many conjectures which suggest that the emergent properties exhibited by
complex automata, resulting from a dynamical  behavior involving
adaptive  mechanisms are very appropriated to describe biological systems.
The use of cellular automata to model the immune responses has been reviewed
recently by one of us \cite{zs98}.

\noindent
{\bf Acknowledgements}: We thank  Nelson Vaz and his group  for the
enlightening discussions. In special we  acknowledge the permission to
reproduce data from reference \cite{ana97} and B. Verdolin  for supplying
 the experimental data shown in figure 1.  We also thank  Luiz  Andrade
for helpful discussions.  ATB acknowledges the kind hospitality of DF-UFMG.
This work was partially supported by the Brazilian
Agencies CNPq, CAPES, FINEP and  FAPEMIG.

\bigskip

\noindent {\bf Figure Captions:}

\noindent{\bf Figure 1:} The specific antibody counts as a function of the 9
 antigen (OVA) presentations (boosters) for a group of $6$ young mice (8 weeks
 old) with one week of interval between them. All data are normalized by
 the first response \cite{bruno97}. In the insert we show the effects of
aging in the specific antibody counts as a function of the antigen
presentation for young (8 weeks old) and old (25 weeks old) mice
\cite{ana97}.

\noindent{\bf Figure 2:} The time evolution  of the difference between the
 Hamming distances
of  ``control''  group and replica group which was subjected to  a sequence of
 antigens
presentation with one week of interval between them. The simulations were
 performed for two different ages: (circle) starting at $t=280$ (8weeks)
and (square) at $t=875$ (25 weeks). In the insert we show the mean 
saturation values for the Hamming distance differences  for the two 
ages considered.

\noindent{\bf Figure 3:} The evolution of the Hamming distance difference
between ``control'' and replica, for one sample, for the damaged region
(DR), the mirror image of the damages region (M) and the overall changes on the
lattice (HD). This simulation corresponds to the immunization process performed
in a $25$ weeks old system, by producing a damage of $6$ concentric layers
($377$ sites). The M region corresponds to 
images and nearest-neighbors of the damaged region.


\begin{references}

\bibitem{Jerne}
Jerne, N.K., {\it Ann. Immuno. (Inst. Pasteur)} {\bf 125 C}, 373-389 (1974).

\bibitem{coutinho}
Coutinho, A., {\it Immunol. Rev.} {\bf 110}, 63-87 (1989).

\bibitem{holmberg}
Holmberg, D., Anderson, \AA , Carlsson, L. \& Forsgren, S.,
{\it Immunol. Rev.} {\bf 110}, 84-103 (1989).

\bibitem{bz97}
Bernardes, A.T. and Zorzenon dos Santos, R.M. {\it J. Theor. Biol.} {\bf 186}
173 (1997).

\bibitem{zb95}
Zorzenon dos Santos, R.M and Bernardes, A.T., {\it Physica A}
{\bf 219}, 1-12 (1995).

\bibitem{sw92}
Stauffer, D. and Weisbuch, G., {\it Physica A} {\bf 180}, 42-52 (1992).

\bibitem{bsp92}
De Boer, R.J., Segel, L.A. and Perelson, A.S.,
{\it J. Theor. Biol.} {\bf 155}, 295-333 (1992).

\bibitem{ps79}
Perelson, A.S. and Oster, G. F.,
{\it J. Theor. Biol.} {\bf 81}, 645-670 (1979).

\bibitem{bruno97}
Verdolin, B., MSc. Thesis, Departamento de Bioqu\'{\i}mica e Imunologia,
Instituto de Ci\^encias Biol\'ogicas, UFMG, Belo Horizonte (1997).

\bibitem{lahmann}
Lahmann, W.M, Menezes, J.S., Verdolin, B.A. and Vaz, N.M., {\it Braz. J. Med.
 Biol. Res.} {\bf 25} 813-821 (1992).

\bibitem{ana97}
Faria, A.M.C., Ficker, S.M., Spezialli, E., Menezes, J.S., Stransky, B.,
 Rodrigues, V.S. and Vaz, N.M.,submitted to {\it Immunology} (1997).

\bibitem{jancar94}
Jan, N. and de Arcangelis,L.,{\it Annual Reviews of Computational
 Physics} {\bf I}, ed. by D. Stauffer,World Scientific, 1-16 (1994).

\bibitem{zs98}
Zorzenon dos Santos, R.M., to appear in {\it Annual Reviews of
Computational Physics} {\bf VI}, ed. by D. Stauffer, World
Scientific (1998).


\end{references}
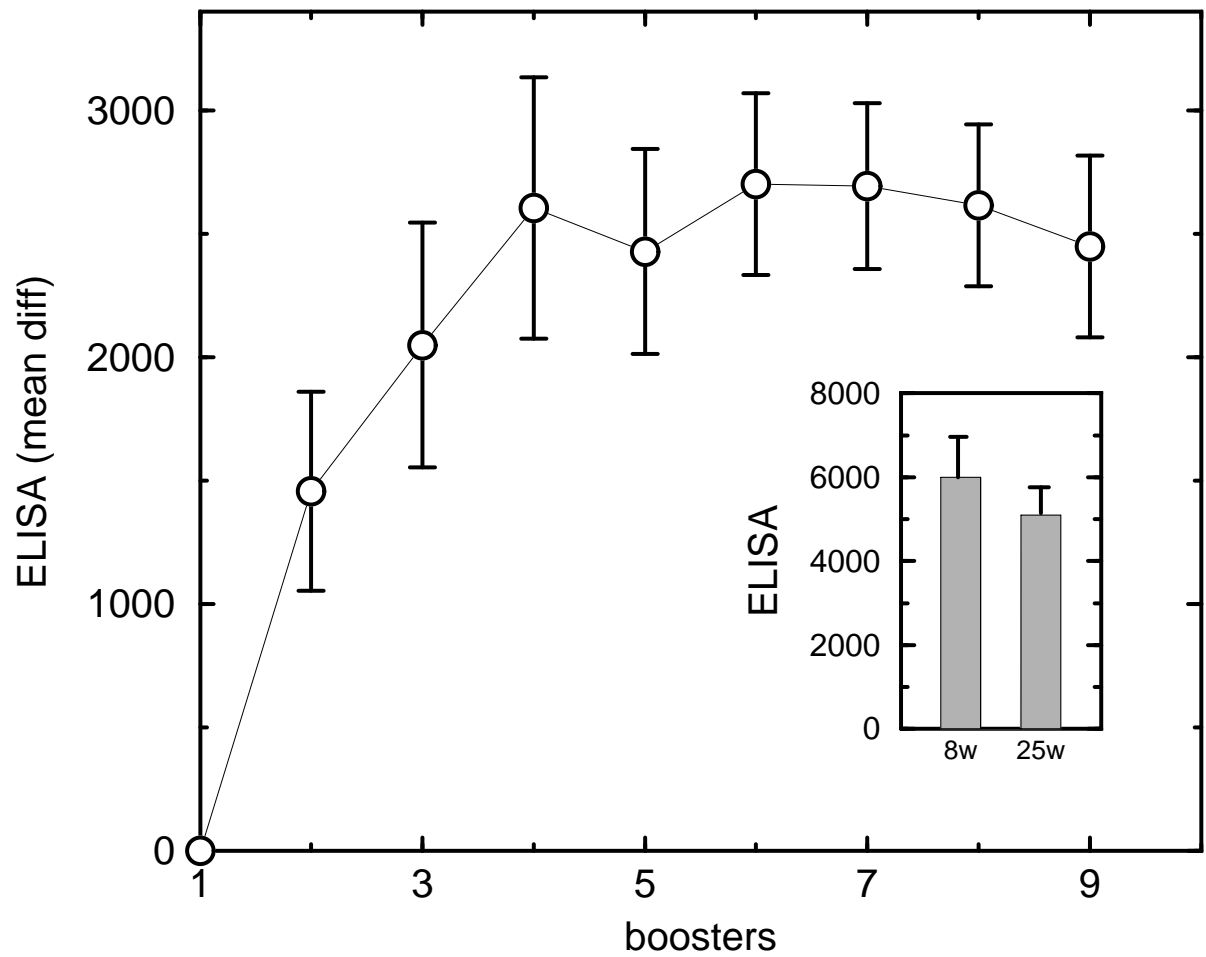
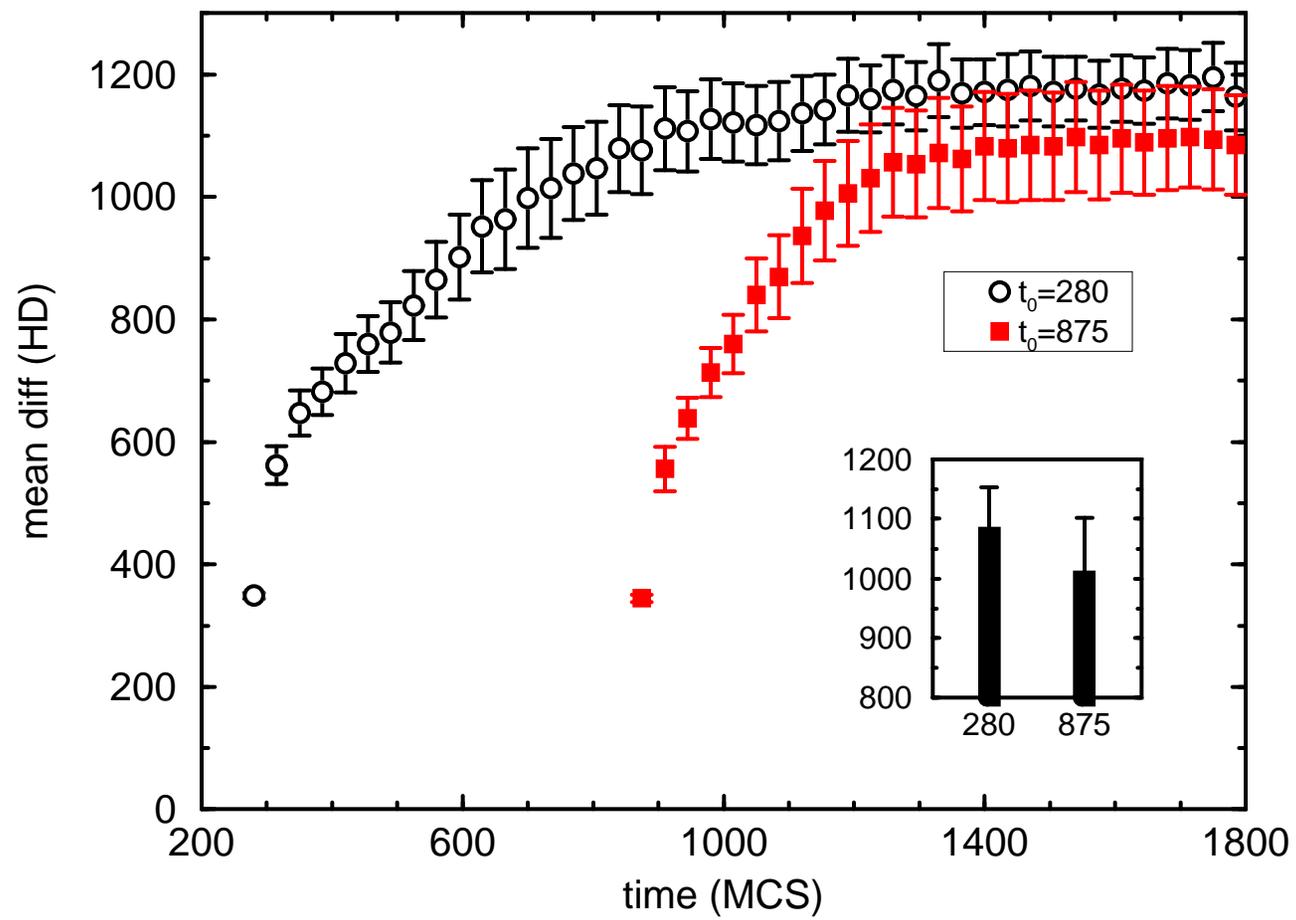
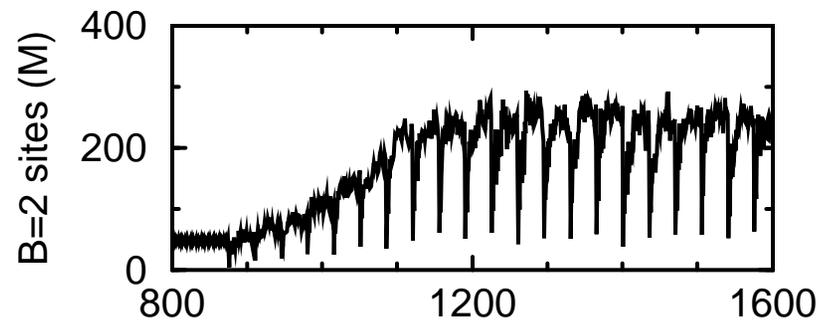
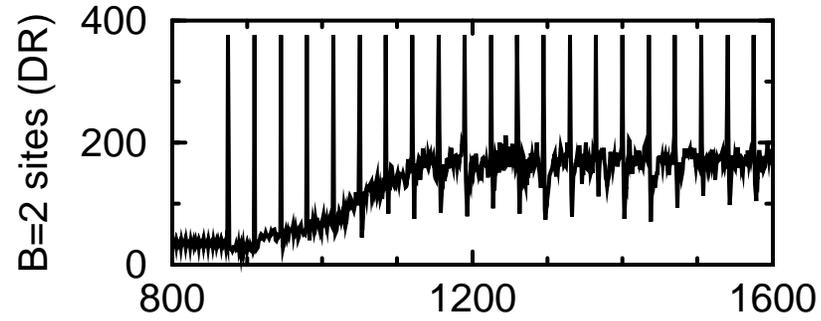
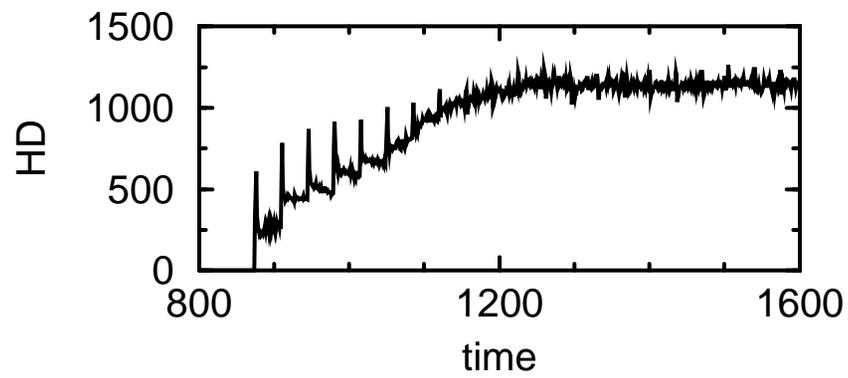
\end{document}